\documentclass[twocolumn,showpacs,preprintnumbers,amsmath,amssymb]{revtex4}
\usepackage{graphicx}
\usepackage{dcolumn}
\usepackage{bm}
\usepackage{natbib}
\usepackage[T1]{fontenc}
\begin{document}

\title{NMR evidence for an inhomogeneous transition between the
ferromagnetic and antiferromagnetic ground states in
Pr$_{1-x}$Ca$_{x}$MnO$_{3}$ manganites}

\author{M.S. Reis}
\altaffiliation[Present address: ]{Centro Brasileiro de Pesquisas
Físicas, Rua Dr. Xavier Sigaud 150 Urca, 22290-180 Rio de
Janeiro-RJ, Brasil} \email{marior@fis.ua.pt}
\author{V.S. Amaral}
\affiliation{Departamento de Física and CICECO, Universidade de
Aveiro, 3810-193 Aveiro, Portugal}
\author{P.B. Tavares}
\affiliation{Departamento de Química, Universidade de
Trás-os-Montes e Alto Douro, 5001-911 Vila Real, Portugal}
\author{A.M. Gomes, A.Y. Takeuchi, A.P. Guimarães and I.S. Oliveira}
\affiliation{ Centro Brasileiro de Pesquisas Físicas, Rua Dr.
Xavier Sigaud 150 Urca, 22290-180 Rio de Janeiro-RJ, Brasil}
\author{P. Panissod}
\affiliation{Institut de Physique et Chimie des Matériaux de
 Strasbourg, UMR 75040 du CNRS, 23 rue du Loess, 67037 Strasbourg, France }%

\date{\today}

\begin{abstract}

The low temperature behavior of Pr$_{1-x}$Ca$_{x}$MnO$_{3}$
manganites are known to undergo a transition at $x\sim$0.3, from
an insulating ferromagnetic state, at low Ca concentration, to an
insulating antiferromagnetic state. Above the onset concentration
for the charge-ordering effect ($x\sim 0.3$), a metal-insulator
transition induced by an external magnetic field is also observed
and related to the collapse of the charge-ordered state. In this
paper we show that the ferro-antiferromagnetic transition takes
place in an inhomogeneous way: around the critical concentration
the sample is a mixture of ferromagnetic and antiferromagnetic
regions. The zero-field NMR measurements show that a fraction of
the ferromagnetic regions is in a fast hopping regime, which
suggests that a small fraction of the sample could be metallic,
even in zero field. This fraction of fast hopping regions is
maximal at $x=$0.3, which is also the concentration for which the
insulator-metal transition has been observed in the smallest
field.

\end{abstract}


\maketitle

\section{Introduction}\label{section_Introduction}

The mixed-valence manganites AMnO$_3$, where $A$ is a trivalent
rare-earth mixed with a divalent alkaline-earth, were the subject
of an enormous quantity of work \cite{PR_344_2001_1}, in
connection with the peculiar magnetic and electrical properties
observed in these systems. Due to the competition of several
interactions (double exchange, superexchange, charge ordering,
Jahn-Teller effect of Mn$^{3+}$, among others), a strong tendency
to phase separation, with a large phase coexistence between the
two phases, appears in these compounds.

In this direction, Allodi \emph{et al.} investigated the
La$_{1-x}$Ca$_{x}$MnO$_3$ \cite{PRL_81_1998_4736},
(LaMn)$_{1-x/6}$O$_3$ \cite{PRB_57_1998_1024}, LaMnO$_{3-x}$
\cite{PRB_56_1997_6036} and Pr$_{0.5}$Sr$_{0.5}$MnO$_3$
\cite{PRB_61_2000_5924} manganites, using the NMR technique. They
stressed out a phase coexistence around the transition, from
ferromagnetic to antiferromagnetic phase, induced by field,
temperature or concentration. In addition, they observed a
peculiar magnetic structure revealed by the NMR spectra, signaling
an inhomogeneous spin arrangement, possibly due to a coexistence
of ferromagnetic and antiferromagnetic clusters. Kapusta \emph{et
al.} \cite{PRL_84_2000_4216}, Kumagai \emph{et al.}
\cite{PRB_59_1999_97} and Dho \emph{et al.}
\cite{PRB_60_1999_14545} enforced the results reported by Allodi
and co-workers.

Particularly interesting is the Pr$_{1-x}$Ca$_x$MnO$_3$ system,
since its phase diagram exhibit a rich variety of magnetic and
electronic structures. Following the complete phase diagram
established by Martin \emph{et al.} \cite{PRB_60_1999_12191}
(figure 1), a ferromagnetic-insulator phase arises for x$<$0.27,
with Curie temperature T$_{C}$ and magnetic moment at 4 K around
100 K and 3.5 $\mu_{\textnormal{\tiny{B}}}$/mol Mn, respectively.
For 0.34$<$x$<$0.85, an antiferromagnetic-insulator phase arises
for temperatures typically below 170 K, coexisting with a
charge-ordering state with onset temperature T$_{CO}$ between 200
K for x=0.35 and 100 K for x=0.85, and an associated
metal-insulator transition induced by magnetic field
\cite{JMMM_200_1999_1,{prb_53_1996_r1689}}. A significant loss of
magnetic moment, comparatively to x$<$0.27, is observed for this
concentration range and this quantity at 4 K reaches 0.01
$\mu_{\textnormal{\tiny{B}}}$/mol Mn. Finally, for 0.27$<$x$<$0.35
a remarkable phase competition between the
antiferromagnetic-insulator charge ordered state and
ferromagnetic-insulator state arises. Note that all values
mentioned above are slightly dependent on the sample preparation
procedure.

Additionally, we can consider the magnetic and electrical effects
due to Ca$^{2+}$-Pr$^{3+}$ substitution as a pure valence effect
\cite{ZPB_104_1997_21}, since the $A$ site ionic radius is kept
constant through the series, i.e., Pr$^{3+}$ and Ca$^{2+}$ ionic
radii are rather similar: 1.179 \AA \: for Pr$^{3+}$ and 1.180 \AA
\: for Ca$^{2+}$ \cite{AC_A32_1976_751}. However, in this series,
even with the A size constant, the volume cell decreases as the
ratio Mn$^{3+}$/Mn$^{4+}$ decreases (increasing $x$), in agreement
with the relation $r_{Mn^{4+}}<r_{Mn^{3+}}$
\cite{ZPB_104_1997_21}.

In the present work we aim to explore the nature of the magnetic
ground state through the concentration range of phase competition,
between the antiferromagnetic-insulator charge ordered state and
the ferromagnetic-insulator state, using, for that, the zero-field
NMR technique.

\section{Experimentals Details}\label{section_Experimental_Details}

The samples were prepared by the ceramic route, starting from the
stoichiometric amount of Pr$_2$O$_3$ (99.9 \%), CaCO$_3$ (>99 \%)
and MnO$_2$ (>99 \%), and heated in air, with five intermediate
crushing/pressing steps. The final crushed powders were compressed
in cylindrical rods, which were sintered in air at 1350
$^\textnormal{o}$C during 45 hours, with a subsequent fast
freezing of the samples. X-ray diffraction patterns confirmed that
the samples are a single orthorhombic phase, with the space group
\textit{Pbnm}. The NMR experiments have been performed on
$^{55}$Mn using a broad band frequency scanning spectrometer with
phase sensitive detection. Spin echo intensity has been recorded
every 1 MHz in the frequency range 170-530 MHz. The spectra were
recorded at several values of the radio-frequency (RF) excitation
field. This procedure allows to measure the NMR enhancement factor
associated with the transverse susceptibility of the electronic
magnetization and to correct the signal amplitude for this
enhancement factor.

\section{Results and Discussion}

Figure 2 displays the spectra observed at 1.4 K for the Ca
concentrations $x=$0.20, 0.25, 0.30 and 0.32. Except for $x=$0.20,
the observed signal is rather low and the integral intensity
(figure 3) decreases by a factor 30 between $x=$0.20 and $x=$0.32.
This means that, in the latter samples, either the NMR enhancement
factor, arising from the transverse magnetic susceptibility, is
small - as in an antiferromagnet - , or that only a part of the
sample is observed. Actually, the enhancement factor measured on
the spectra is in the range 50 to 100, which shows that the signal
arises from ferromagnetic domains. Thus, no signal from
antiferromagnetic regions could be observed at the maximum
available RF power.

In previous works \cite{EPJB_11_1999_243,{PRB_62_2000_9532}} the
existence of a small residual ferromagnetic component above
$x=$0.3 has been assigned to a possibility of a small canting of
the moments in the antiferromagnetic state. In such a case, all
the Mn nuclei would have been observed and the decrease of the NMR
signal would have been due to the decrease of the enhancement
factor associated with the decrease of the effective ferromagnetic
moment. The present measurements show that the enhancement factor
does not decrease. It even increases from 50 to 100 between
$x=$0.20 and $x=$0.32. Therefore, we can unambiguously conclude
that the samples always contains some ferromagnetic parts and that
the small residual moment in the antiferromagnetic state is due to
the inhomogeneous character of the ferro-antiferromagnetic
transition.

Assuming that the whole sample is ferromagnetic for $x=$0.20, the
measured intensity for $x=$0.32 shows that a volume fraction of
about 3{\%} of the sample is still ferromagnetic, in agreement
with the residual ferromagnetic component measured on standard
magnetization curves. A detailed analysis of the magnetic and
thermal properties of this Pr$_{1-x}$Ca$_{x}$MnO$_3$ manganites is
discussed elsewhere \cite{magnetocalorico_prca}.

Whether the magnetic inhomogeneity of the samples is intrinsic or
extrinsic is a matter of discussion. Indeed, it can be speculated
that the residual ferromagnetic fraction in the antiferromagnetic
phase arises from the presence of defects and/or impurities, but,
concentration dependence? However, the intrinsic fluctuations of
composition in a random alloy can also explain the observation as
shown below.

Let us assume that $d_{m}$ is the minimum diameter for a
statistical cluster to show ferromagnetic or antiferromagnetic
order. In other words, $d_{m}$ is a typical coherence length for
the magnetic order. Be $n$ the number of formula units within a
sphere of diameter d$_{m}$, and, on average, such spheres contain
$m=nx$ Ca atoms. Let us now assume that $x_{c}$ is the critical
local concentration, above (below) which the sphere shows an
antiferromagnetic (ferromagnetic) coherence. Correspondingly,
$m_{c}=nx_{c}$ is the critical number of Ca atoms within the
sphere above (below) which the sphere is antiferromagnetic
(ferromagnetic).

However, there are statistical concentration fluctuations between
the spheres. For a sample with nominal concentration $x$, the
probability $P_{x,n}(k)$ to find $k$ Ca atoms inside a sphere of
diameter d$_{m}$, is given by the binomial law:

\begin{equation} \label{binomial_distribution}
P_{x,n}(k) = \frac{n!}{k!(n-k)!}x^k(1 - x)^{n-k}
\end{equation}
and the probability $P_{f}(x)$ to find a ferromagnetic sphere is:

\begin{equation}\label{probabilidade_esfera_ferro}
P_f(x) = \sum\limits_{k=0}^{m_c}P_{x,n}(k)
\end{equation}

Considering that $n$ is most probably large, a continuous
approximation for the binomial probability can be safely used in
the form of the normal distribution:

\begin{equation}\label{normal_distribution}
P_{x,n}(y)=\frac{1}{\sqrt{2\pi[x(1-x)]/n}}\exp(-\frac{(y-x)^2}{2[x(1-x)]/n})
\end{equation}
where $y=k/n$ is the local concentration within the sphere. The
average value $\langle y \rangle=x$ and variance $\sigma
^2=x(1-x)/n$ of the normal distribution takes the same values as
those of the binomial law.

Then, the fraction of ferromagnetic phase in a sample with nominal
Ca concentration $x$ is :

\begin{equation}\label{ff}
F_f(x)=\int\limits_0^{x_c } P_{x,n}(y)dy
\end{equation}

\begin{eqnarray}\label{sigmoidal_solved}
F_f(x)=\frac{1}{2} \left\{\
\textnormal{erf}\left[\sqrt{\frac{n}{2x(1-x)}}(x_c-x)\right]
\right. + \nonumber \\
\left. \textnormal{erf}\left[\sqrt{\frac{n}{2x(1-x)}}x\right]
\right\}
\end{eqnarray}
where $\textit{erf}$ is the \emph{error function}, with $x_c$ and
$n$ acting as free parameters. This function $F_f(x)$ is centered
on $x_{c}$, with a width $\sqrt{x_{c}(1-x_{c})/n}$. Although we
have not enough data points to make an accurate fit, we can
estimate that the minimum number of formula units contained in
ferromagnetic clusters is of the order of $n=$150. Additionally,
the critical local concentration $x_c=0.24$ above (below) which an
antiferromagnetic (ferromagnetic) order is established, was also
found. Figure 3 presents the normalized spectra integral
intensity, considering that $x=$0.20 is fully ferromagnetic. The
solid line is calculated from Eq.\ref{sigmoidal_solved}.

As to the spectral shapes, they are nearly identical at both ends
of the studied concentration range ($x=$0.20 and $x=$0.32). They
show a narrow line at 320 MHz and a broad structured hump in the
370-440 MHz range. As the samples are insulators, one indeed
expects two lines arising from the $^{55}$Mn nuclei: a line from
the Mn$^{4+}$ sites around 300 MHz, and a line from the Mn$^{3+}$
sites around 400 MHz. The broadening of the upper frequency line
is understood as arising from the Jahn-Teller distortion around
the Mn$^{3+}$ sites, which induces magnetic and electric
anisotropies. However, the observed intensity ratio of the two
lines is not that expected from the charge balance: an intensity
ratio of x/(1-x) is expected between lower (Mn$^{4+}$) and upper
(Mn$^{3+}$) frequency  lines, whereas the observed intensities are
about equal. The reason for the discrepancy has to be found in the
existence of a $^{141}$Pr line around 320 MHz. Indeed, the
$^{141}$Pr resonance has been identified in this low frequency
range in a previous study of the Pr$_{1-x}$Ca$_x$MnO$_{3}$ system
\cite{PRB_62_2000_9532}. Then, the expected ratio between the
lower (Pr$^{3+}$ and Mn$^{4+})$ and upper (Mn$^{3+}$) frequency
lines is 1/(1-x), in much better agreement with the observation.
Figure 4 shows a fit of the spectrum observed for $x=$0.20, using
two gaussian lines for Pr and Mn$^{4 + }$around 320 MHz, and a
quadrupolar split spectrum including magnetic anisotropy for
Mn$^{3+}$, centered at 400 MHz. The relative integral intensities
of the three lines are in the ratio 0.84/0.21/0.75, in rather fair
agreement with the (1-x)/x/(1-x) ratio expected from the charge
balance.

As mentioned above, the spectrum observed for $x=$0.32 is nearly
identical to the one observed for $x=$0.20, even though its
integral intensity is 30 times smaller. This means that the
residual ferromagnetic regions have a Ca content close to 20 \%.
The slight change of shape of the Mn$^{3 + }$ line (380-420 MHz)
can be explained, from the simulation we have made, by a rotation
of 90\r{ } of the easy axis with respect to that in the totally
ferromagnetic sample.

By contrast, the spectra for the intermediate compositions are
significantly different, as they show a significant intensity
peaking in the intermediate frequency range 360-380 MHz (see
figure 2). $^{55}$Mn NMR has been observed in this range in
metallic manganites \cite{MSEB_63_1999_22}. Indeed, in samples
where the hopping rate is fast (faster than the NMR frequency), it
averages the hyperfine field for the two ionic configurations,
resulting in a single line at the center of gravity of the two
$^{55}$Mn lines observed in insulating samples. This mere
observation shows that the two samples with $x=$0.25 and $x=$0.30
embed regions where the hopping rate is fast, even in zero field.
The fast hopping signal is maximum for $x=$0.30, which is in
agreement with the fact that this sample shows also the lowest
field for the insulator-metal transition \cite{JMMM_200_1999_1}.
Whether the fast hopping regions are metallic or not is still an
open question. Considering the intensity of the fast hopping line,
which is at most 1$/$3 of the total intensity, one can deduce that
it corresponds to a volume fraction of 5{\%} for $x=$0.30. This
would be too small a fraction to make the whole sample conducting.

\section{Conclusion}

There are already a number of experimental evidences for the
coexistence of two phases around the various transitions exhibited
by the mixed-valency manganites systems (see
Ref.\cite{PR_344_2001_1} and references therein). The present
study has shown that this is also the case for the transition
between the insulating-ferromagnetic state and the
insulating-antiferromagnetic state in Pr$_{1-x}$Ca$_{x}$MnO$_3$.

Although the extrinsic nature of the phase admixture cannot be
excluded, we have shown that a statistical model that takes into
account the concentration fluctuations in a random homogeneous
alloy, can explain the observed results. The volume fraction of
the ferromagnetic phase observed by NMR agrees with that
calculated within the model provided here, where a ferromagnetic
(antiferromagnetic) order establishes in a statistical cluster
with a local Ca concentration smaller (larger) than $x=$0.24. The
minimum coherent size of such clusters would correspond to about
150 unit cells.

In addition, we have evidenced, in a limited concentration range
of fast hopping, possible metallic regions in a virgin samples and
zero external magnetic field. The volume fraction of these
"metallic" regions is, at most, 5 \%. The maximum fraction is
observed for $x=$0.30, which is also the concentration that
requires the smaller external magnetic field to induce the
insulator-metal transition.

\section{Acknowledgements}

The authors thanks FAPERJ and CAPES/Brasil and FCT/Portugal
(contract POCTI/CTM/35462/99), for financial support, and P.P. the
support of the CNPq/CNRS cooperation.

\section{Figure Captions}

Figure 1 - Magnetic and electrical phase diagram for
Pr$_{1-x}$Ca$_{x}$MnO$_3$. After Martin \emph{et al.}
\cite{PRB_60_1999_12191}.

Figure 2 - $^{55}$Mn zero field NMR spectra in
Pr$_{1-x}$Ca$_{x}$MnO$_3$ at 1.4 K, for $x=$0.20, 0.25, 0.30 and
0.32. This range of concentration cover the region of the phase
diagram where ferro and antiferromagnetic order coexist.

Figure 3 - NMR integral intensity vs. concentration. The closed
circles represent the ferromagnetic fraction of each sample
obtained from the integral intensity of spectra, whereas the solid
line represent our statistical model for concentration
fluctuations in a random homogeneous alloy
(Eq.\ref{sigmoidal_solved}).

Figure 4 - Zero field NMR spectra at 1.4 K, for $x=$0.20. The
solid lines represent the Gaussian fit of each resonance peak.

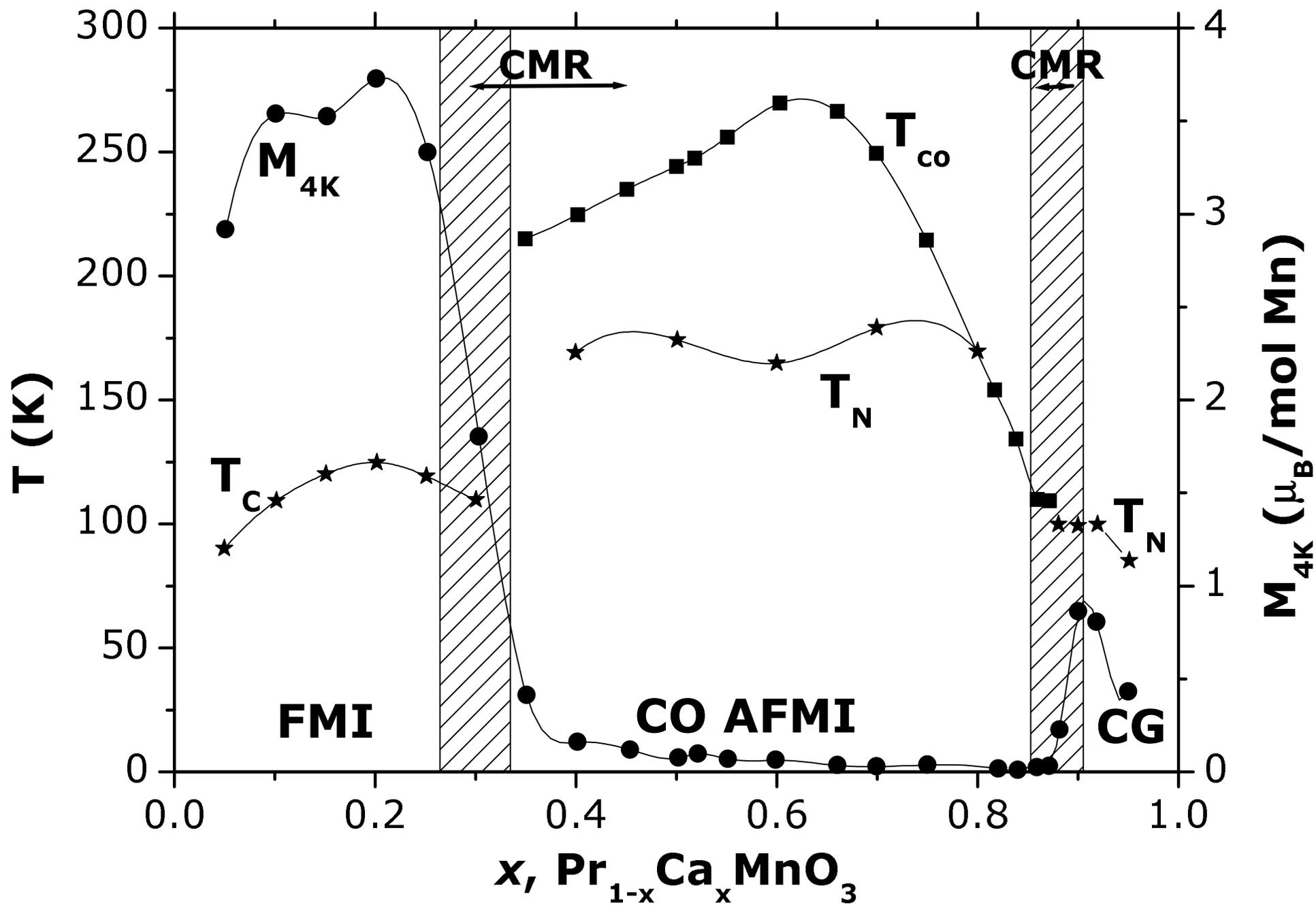
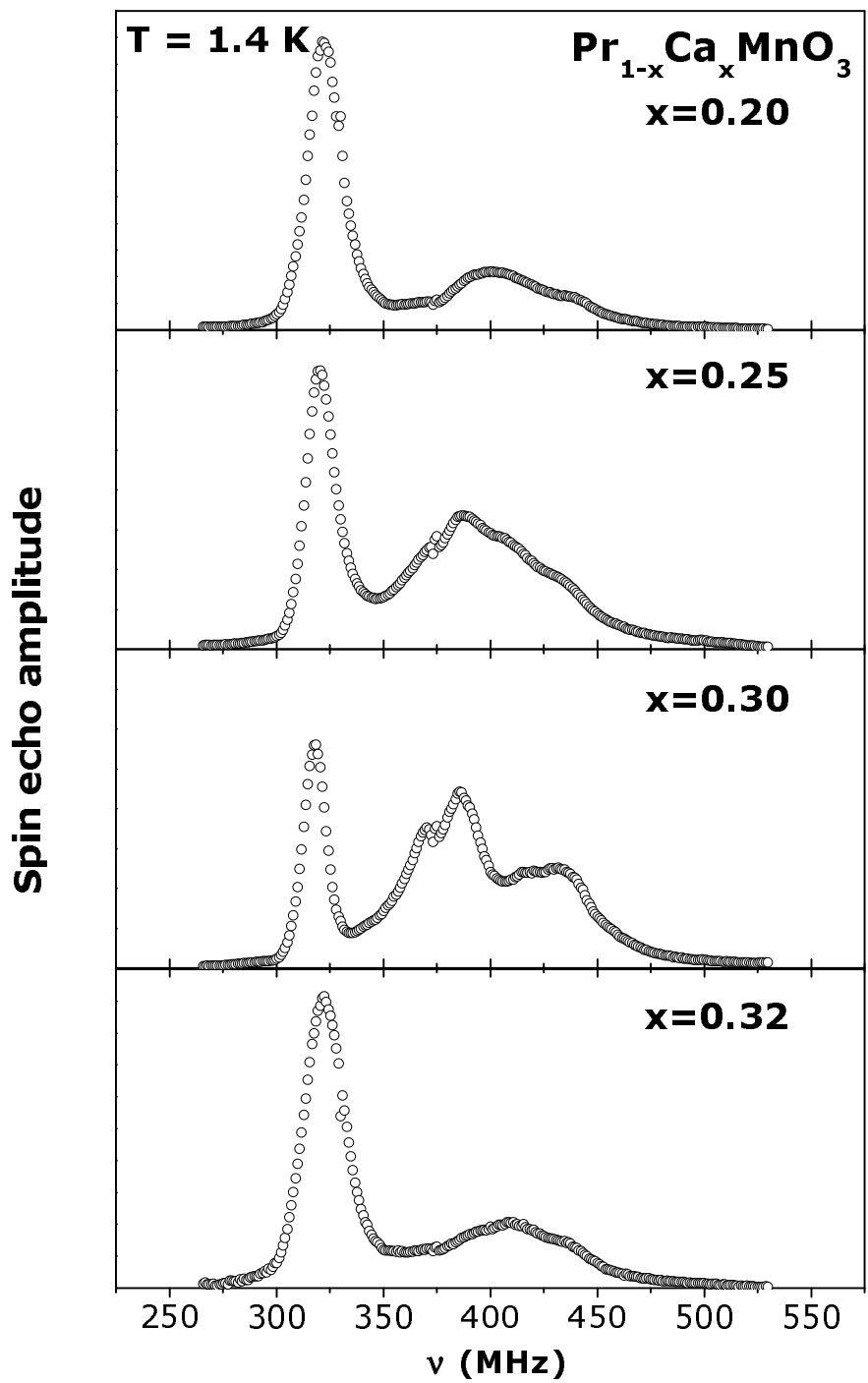
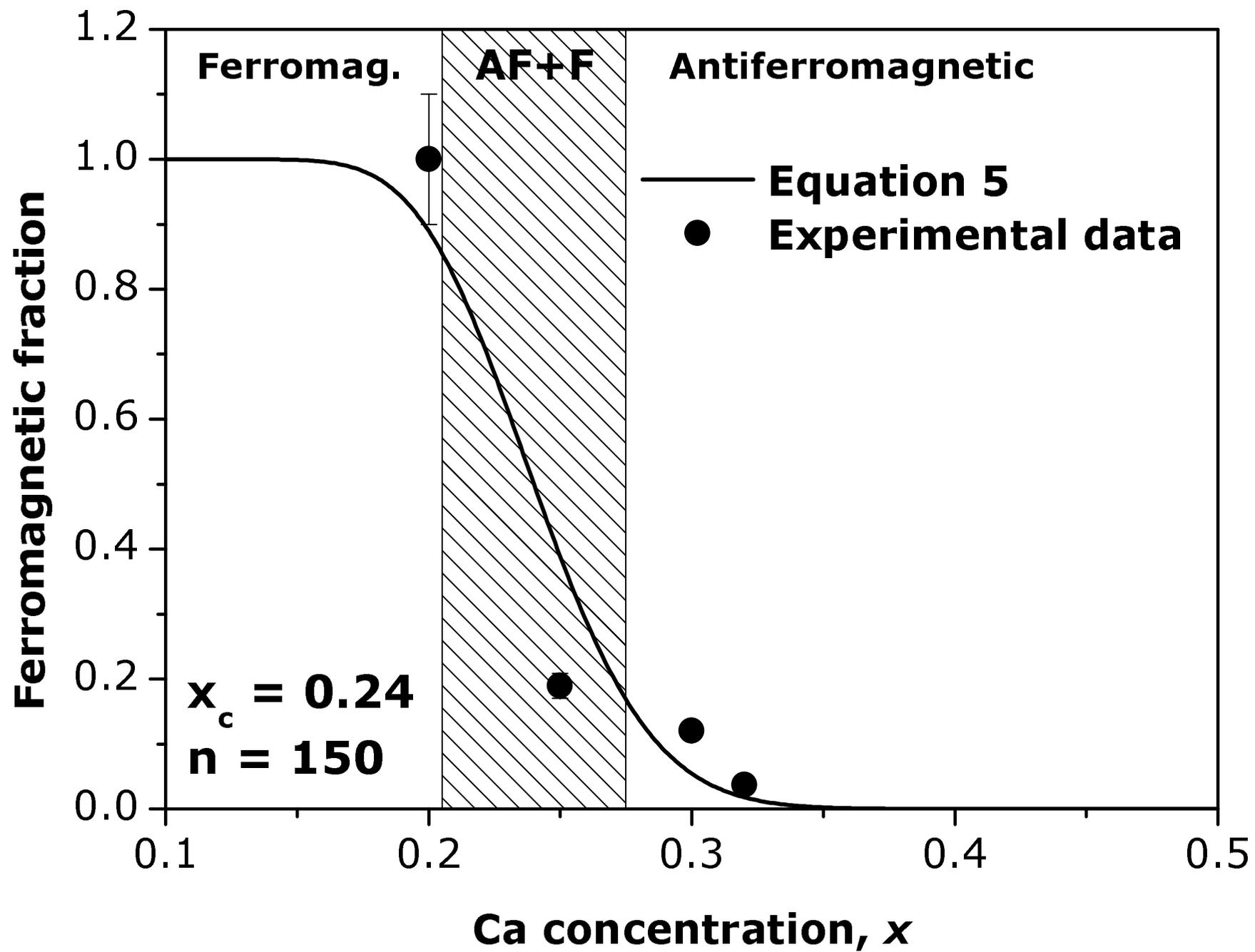
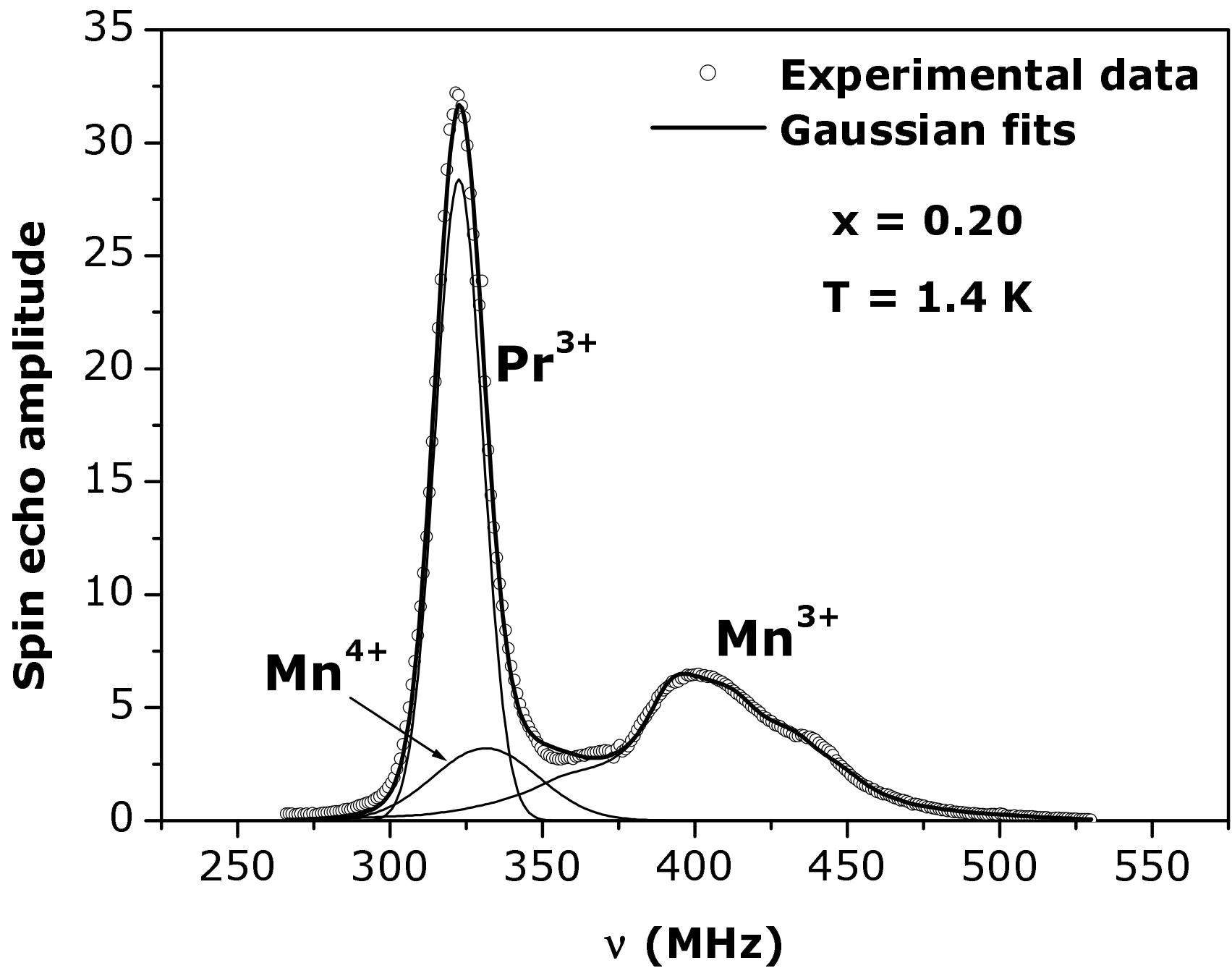
\end{document}